\RequirePackage{fix-cm}
\documentclass[smallextended]{svjour3}       
\smartqed  
\usepackage{bm} 
\usepackage{graphicx,color}
\usepackage{url}
\newcommand{\fadd}[1]{#1}
\newcommand{\fdel}[1]{}
\newcommand{\finfo}[1]{}
\newcommand{\ffadd}[1]{#1}
\newcommand{\ffinfo}[1]{}
\newcommand{\hadd}[1]{#1}
\newcommand{\yadd}[1]{#1}
\newcommand{\yyadd}[1]{#1}
\newcommand{\yyyadd}[1]{#1}
\begin{document}

\title{EigenKernel
\thanks{The present research was partially supported by JST-CREST
project of 'Development of an Eigen-Supercomputing Engine 
using a Post-Petascale Hierarchical Model',  
Priority Issue 7 of the post-K project and KAKENHI funds (16KT0016,17H02828).
Oakforest-PACS was used through the JHPCN Project (jh170058-NAHI)
and through Interdisciplinary Computational Science Program in Center 
for Computational Sciences, University of Tsukuba.
The K computer was used in
the HPCI System Research Projects (hp170147, hp170274, hp180079, hp180219).
Several computations were carried out also on the facilities of the Supercomputer Center, the Institute for Solid State Physics, the University of Tokyo.
}
}
\subtitle{A middleware for parallel generalized eigenvalue solvers to attain high scalability and usability}


\author{Kazuyuki Tanaka \and
        Hiroto Imachi  \and Tomoya Fukumoto \and Akiyoshi Kuwata \and Yuki Harada  \and Takeshi Fukaya \and Yusaku Yamamoto \and Takeo Hoshi 
}


\institute{K. Tanaka T. Fukumoto A. Kuwata Y. Harada T. Hoshi \at
              Department of Applied Mathematics and Physics , Tottori University, 4-101 Koyama-Minami, Tottori, 680-8552, Japan  \\
              Tel.: +81-857-31-5448 \\
              Fax: +81-857-31-5747 \\
              \email{hoshi@damp.tottori-u.ac.jp}           
           \and
           H. Imachi \at
              Department of Applied Mathematics and Physics , Tottori University, 4-101 Koyama-Minami, Tottori, 680-8552, Japan \\
               \emph{Present address:  Preferred Networks, Inc.}   
           \and
           T. Fukaya \at
              Information Initiative Center, Hokkaido University, Japan \\
           \and
           Y. Yamamoto \at
              \fadd{Department of Communication Engineering and Informatics,} 
              The University of Electro-Communications, Japan \\
 }

\date{Received: date / Accepted: date}

\maketitle

\begin{abstract}
An open-source middleware EigenKernel 
was developed 
for use with parallel generalized eigenvalue solvers
or large-scale electronic state calculation 
to attain high scalability and usability\fdel{(\verb|https://github.com/eigenkernel/|)}.
\finfo{removed URL}
The middleware enables the users to choose
the optimal solver,
among the three parallel eigenvalue libraries
of ScaLAPACK, ELPA, EigenExa and hybrid solvers constructed from them, 
according to the problem specification and the target architecture.
The benchmark was carried out on 
the Oakforest-PACS supercomputer and 
reveals that 
ELPA, EigenExa and their hybrid solvers show better performance,
when compared with pure ScaLAPACK solvers.
The benchmark on the K computer is also used for discussion.
In addition,
a preliminary research for the performance prediction was investigated, 
so as to predict the elapsed time $T$ 
as the function of the number of used nodes $P$ ($T=T(P)$).
The prediction is based on Bayesian inference \yyyadd{using} the Markov Chain Monte Carlo (MCMC) method
and the test calculation indicates that the method is applicable 
not only to performance interpolation but also to extrapolation. 
Such a middleware is of crucial importance 
for application-algorithm-architecture co-design
among the current, next-generation (exascale), and future-generation (post-Moore era) supercomputers.
\end{abstract}

\begin{figure*}
  \includegraphics[width=0.95\textwidth]{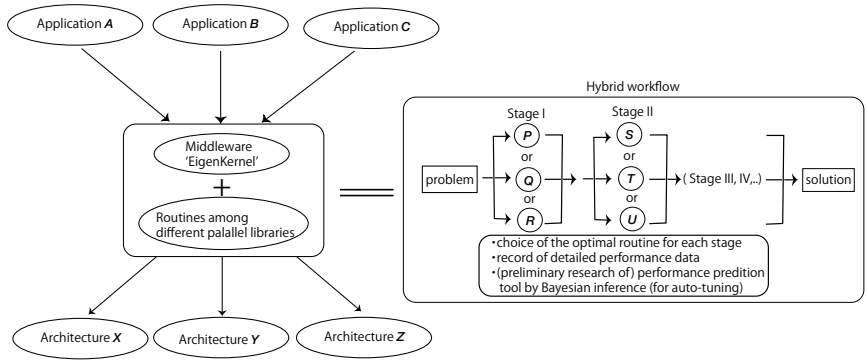}
\caption{Schematic figure of the middleware approach that realizes a hybrid workflow. 
The applications, such as electronic state calculation codes, are denoted
as $A,B,C$ and the architectures (supercomputers) are denoted as $X,Y,Z$. 
A hybrid workflow for a numerical problem, such as generalized eigenvalue problem,
consists of Stages I, II, III... and
one can choose the optimal one for each stage
among the routines $P, Q, R, S, T, U$ 
in different numerical libraries. 
}
\label{FIG_CONCEPT}       
\end{figure*}

\section{Introduction \label{SEC-INTRO}}

Efficient computation with
the current and 
\fadd{upcoming (both exascale and post-Moore era) supercomputers}
can be realized by application-algorithm-architecture co-design~\fadd{\cite{Shalf11,Dosanjh14,POST-K,CoDEx,EuroExa}},
in which various numerical algorithms should be prepared
and the optimal one should be chosen 
according to the target application, architecture and problem. 
For example, an algorithm designed to minimize the floating-point operation count can be
the fastest for some combination of application and architecture,
while another algorithm designed to minimize communications 
(e.g. the number of communications or the amount of data moved)
can be the fastest in another situation.

The present paper proposes a middleware approach,
so as to choose the optimal set of numerical routines  
for the target application and architecture.
The approach is shown schematically in Fig.~\ref{FIG_CONCEPT}
and the crucial concept is called \lq hybrid solver'. 
In general, a numerical problem solver in simulations 
is complicated and consists of sequential stages, 
as Stages I, II, III,... in Fig.~\ref{FIG_CONCEPT}.
Here the routines of $P, Q, R$ are considered for Stage I and
those of $S, T, U$ are for Stage II. 
The routines in a stage \yadd{are} equivalent in their input and output quantities 
but use different algorithms. 
The routines are assumed to be included in 
ScaLAPACK and other parallel libraries. 
Consequently, they show different performance characteristics and the optimal routine
depends not only on the applications denoted as $A, B, C$ but also on
the architectures denoted as $X, Y, Z$. 
Our middleware assists the user to choose the optimal routine among different libraries 
for each stage and such a workflow is called \lq hybrid workflow'. 
The present approach for hybrid workflow is realized by the following functions\yyyadd{.}
First, it provides a unified interface to the solver routines.
In general, different solvers have different user interface, such as the
matrix distribution scheme, so the user is often required to rewrite the 
application program to switch from one solver to another.
Our middleware absorbs this difference and frees the user from this troublesome task.
Second, it outputs detailed performance data such as the elapsed time of each
routine composing the solver. Such data will be useful for detecting the performance
bottleneck and finding causes of it, as will be illustrated in this paper.
In addition, we also \yadd{focus} on a performance prediction function, which predicts the elapsed time of the
solver routines from existing benchmark data prior to actual computations.
As a preliminary research, 
such a prediction method is constructed with Bayesian inference in this paper.  
Performance prediction will be valuable for choosing an appropriate job class in the case of
batch execution, or choosing an optimal number of computational nodes that can be
used efficiently without performance saturation. 
Moreover, performance prediction \yadd{will form the basis of an auto-tuning function planned for the future version, which obviates the need to} care about the job class and detailed calculation conditions.
In this way, our middleware is expected to enhance the usability of existing
solver routines and allows the users to concentrate on computational science itself.


Here we focus on a middleware for the generalized eigenvalue problem (GEP) with real-symmetric coefficient matrices, 
since GEP forms the numerical foundation of electronic state calculations.
Some of the authors developed a prototype of such middleware on the K computer 
in 2015-2016 \cite{IMACHI-JIT2016,HOSHI2016SC16}.
After that, the code appeared at GITHUB 
as EigenKernel ver.~2017  \cite{EIGENKERNEL-URL}
under the MIT license. 
It was confirmed that EigenKernel ver.~2017 works well also on
Oakleaf-FX10 and Xeon-based supercomputers \cite{IMACHI-JIT2016}.
In 2018, a new version of EigenKernel was developed and appeared on the developer branch at GITHUB. 
This version can run on 
\fadd{the} Oakforest-PACS \fdel{(OFP)}\ffadd{supercomputer}, 
a new supercomputer \fdel{using}\fadd{equipping} Intel Xeon Phi many-core processor\fadd{s}. 
In this paper, we take up this version. 
A related project is 
ELSI (ELectronic Structure Infrastructure) that 
provides interfaces to various numerical methods to solve or circumvent GEP in electronic structure calculations
\cite{ELSI-URL,ELSI-PAPER}. 
The present approach limits the discussion to GEP solver and enables 
the user to construct a {\it hybrid} workflow which combines routines from different libraries, 
as shown in Fig.~\ref{FIG_CONCEPT},
while ELSI allows the user to choose a library only {\it as a whole}.  
The present approach of hybrid solver 
will add more flexibility and increase the chance to get higher performance.

This paper presents two topics. First, we show the performance data of various GEP solvers on \fdel{OFP}\fadd{Oakforest-PACS} obtained using EigenKernel. Such data will be of interest on its own since \fdel{OFP}\fadd{Oakforest-PACS} is a new machine and few performance results of \ffadd{dense} matrix solvers on it have been reported; 
\ffadd{stencil-based application \cite{Hirokawa18} and communication-avoiding iterative solver for a sparse linear system \cite{Idomura17} were evaluated on Oakforest-PACS, but their characteristics are totally different from those of dense matrix solvers such as GEP solvers.}
Furthermore, we point out that one of the solvers has a severe scalability problem and investigate the cause of it with the help of the detailed performance data output by EigenKernel. This illustrates how EigenKernel can be used effectively for performance analysis. Second, we describe the new performance prediction method implemented as a Python program. It uses Bayesian inference and predicts the execution time of a specified GEP solver as a function of the number of computational nodes. We present the details of the mathematical performance models used in it and give several examples of performance prediction results. It is to be noted that our performance prediction method can be used not only for interpolation but also for extrapolation, that is, for predicting the execution time at a larger number of nodes from the results at a smaller number of nodes. There is a strong need for such prediction among application users.

This paper is organized as follows.
The algorithm and features of EigenKernel are described in Sec.~\ref{SEC-EIGENKERNEL}. 
Sec.~\ref{SEC-PERFORMANCE-ANA} is devoted to the scalability analysis of various GEP solvers on \fdel{OFP}\fadd{Oakforest-PACS}, which was made possible with the use of EigenKernel. Sec.~\ref {SEC-PREDICTION} explains our new performance prediction method, focusing on the performance models used in it and the performance prediction results in the case of extrapolation. Sec.~\ref{DISCUSSION} discusses our performance prediction method, comparing it with some existing studies. Finally Sec.~\ref{SUMMARY} provides summary of this study and some future outlook.

\section{EigenKernel \label{SEC-EIGENKERNEL}}

EigenKernel is a middleware for GEP that enables the user to use optimal solver routines according to the problem specification (matrix size, etc.) and the target architecture. In this section, we first review the algorithm for solving GEP and describe the solver routines adopted by EigenKernel. Features of EigenKernel are also discussed.

\subsection{Generalized eigenvalue problem and its solution}

We consider the generalized eigenvalue problem 
\begin{eqnarray}
A \bm{y}_k = \lambda_k B \bm{y}_k, 
\label{EQ-GEP-ORG}
\end{eqnarray}
where the matrices $A$ and $B$ are $M \times M$ real symmetric ones 
and $B$ is positive definite ($B \ne I$).
The $k$-th eigenvalue or eigenvector is 
denoted as $\lambda_k$ or $\bm{y}_k$, respectively ($k=1,2,...,M$).
The algorithm to solve Eq.~(\ref{EQ-GEP-ORG}) proceeds as follows. First, the Cholesky decomposition of $B$ is computed, producing an upper triangle matrix $U$ that satisfies
\begin{eqnarray}
B = U^{\rm T} U.
\label{EQ-CHOL}
\end{eqnarray}
Then the problem is 
reduced \yyyadd{to} a standard eigenvalue problem (SEP) 
\begin{eqnarray}
A' \bm{z}_k = \lambda_k \bm{z}_k 
\label{EQ-SEP-ORG}
\end{eqnarray}
with the real-symmetric matrix of 
\begin{eqnarray}
A' = U^{\rm -T} A U^{-1}.
\label{EQ-GEN-MAT-A2}
\end{eqnarray}
When the SEP of Eq.~(\ref{EQ-SEP-ORG}) is solved, 
the eigenvector of the GEP is obtained by 
\begin{eqnarray}
\bm{y}_k = U^{-1} \bm{z}_k.
\label{EQ-BACK}
\end{eqnarray}

The above explanation indicates that
the whole solver procedure can be decomposed into the two parts of
(I) the SEP solver of Eq.~(\ref{EQ-SEP-ORG}) and
(II) the \lq reducer' or 
the reduction procedure between GEP and SEP
by Eqs.~(\ref{EQ-CHOL})(\ref{EQ-GEN-MAT-A2})(\ref{EQ-BACK}).

\begin{figure*}
  \includegraphics[width=0.70\textwidth]{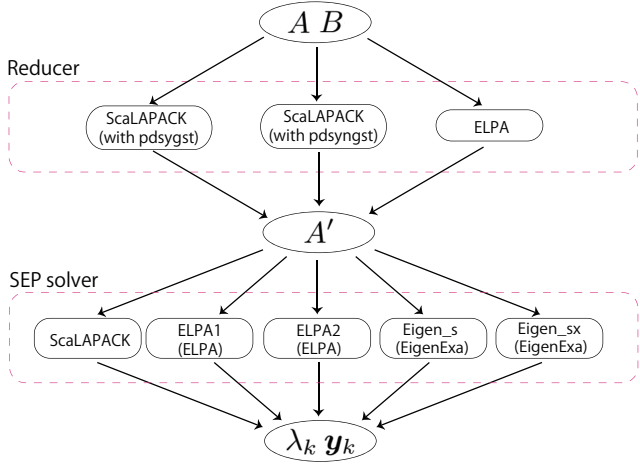}
\caption{Possible workflows for GEP solver. }
\label{FIG-WORKFLOW}       
\end{figure*}

\subsection{GEP Solvers and hybrid workflows \label{SEC-GEP-HYB}}

EigenKernel builds upon three parallel libraries for GEP: ScaLAPACK \cite{SCALAPACK}, ELPA \cite{ELPAWEB} and EigenExa \cite{EigenExaWeb}. 
\ffinfo{changed reference of ELPA and EigenExa: paper $\rightarrow$ web site like ScaLAPACK}
Reflecting the structure of the GEP algorithm stated above, all of the GEP solvers from these libraries consist of two routines, namely, the SEP solver and the reducer. EigenKernel allows the user to select the SEP solver from one library and the reducer from another library, by providing appropriate data format/distribution conversion routines. 
\ffadd{We call the combination of an SEP solver and a reducer a {\it hybrid workflow}, or simply {\it workflow}}. Hybrid workflows enable the user to attain maximum performance by choosing the optimal SEP solver and reducer independently.

Among the three libraries adopted by EigenKernel, ScaLAPACK is the {\it de facto} standard parallel numerical library. However, it was developed mainly in 1990's 
and thus some of its routines show severe bottlenecks on current supercomputers. 
Novel solver libraries of ELPA and EigenExa were proposed, 
so as to overcome the bottlenecks in eigenvalue problems. 
EigenKernel v.2017 were developed mainly in 2015-2016, so as to realize
hybrid workflows among the three libraries. 
The ELPA code was developed in Europe under the tight-collaboration between computer scientists and material science researchers and its main target application is 
FHI-aims (\fdel{=}\finfo{removed $=$}Fritz Haber Institute {\it ab initio} molecular simulations package)
\cite{FHI-AIMS}, a famous electronic state calculation code. 
The EigenExa code, on the other hand, was developed at RIKEN in Japan. 
It is an important fact that
the ELPA code has routines optimized for X86, IBM BlueGene and AMD architectures \cite{ELPA},
while the  EigenExa code was developed so as to be optimal mainly on the K computer. 
The above fact motivates us to develop a hybrid solver workflow 
so that we can achieve optimal performance for any problem on any architecture.
EigenKernel supports only limited versions of ELPA and EigenExa, 
since the update of ELPA or EigenExa requires us, sometimes,  
to modify the interface routine without backward compatibility.
EigenKernel v.2017 supports ELPA 2014.06.001 and EigenExa 2.3c.
In 2018, EigenKernel was updated in the developer branch on GITHUB and can run on Oakforest-PACS.
The benchmark on Oakforest-PACS in this paper
was carried out by the code with the commit ID of 373fb83 
that appeared at Feb 28, 2018 on GITHUB, except where indicated.   
The code is called the \lq current'  code hereafter
and supports ELPA v. 2017.05.003 and EigenExa 2.4p1.

Figure \ref{FIG-WORKFLOW} shows the possible workflows in EigenKernel. 
The reducer can be chosen from two ScaLAPACK routines and the ELPA-style routine
\fadd{,} and the difference between them is discussed later in this paper.  
The SEP solver for Eq.~(\ref{EQ-SEP-ORG}) can be chosen from 
the five routines:
\fdel{a}\fadd{the} ScaLAPACK routine denoted as ScaLAPACK, two ELPA routines denoted as ELPA1 and ELPA2 and two EigenExa routines denoted as Eigen\_s and Eigen\_sx.
The ELPA1 and Eigen\_s routines are based on
the conventional tridiagonalization algorithm like the ScaLAPACK routine
but are different in their implementations.
The ELPA2 and Eigen\_sx routines are based on
non-conventional algorithms for modern architectures. 
Details of these algorithms can be found in the references \fdel{\cite{ELPA,EIGENEXA}}
\fadd{(see \cite{ELPA11,ELPA,ELPAWEB} for ELPA and 
\cite{Imamura11,EIGENEXA,FukayaPDSEC15,EigenExaWeb} for EigenExa)}.

EigenKernel focuses on 
the eight solver workflows
for GEP, which are listed  
as $A, A2, B, C, D, E, F, G$ in Table \ref{TABLE-WORKFLOW}. 
The algorithms of the workflows in Table \ref{TABLE-WORKFLOW} 
are explained in our previous paper 
\cite{IMACHI-JIT2016}, except the workflow $A2$.  
The workflow $A2$ is quite similar to \yyyadd{$A$} and the difference between them
is only the point that the ScaLAPACK routine pdsyngst, one of the reducer routines, 
is used in the workflow $A2$,  
instead of pdsygst in the workflow $A$.
\yyadd{The pdsygst routine is a distributed parallel version of the dsygst routine in LAPACK.} 
\yyadd{This routine repeatedly calls the triangular solver, namely pdtrsm, with a few right-hand sides, and this part often becomes a serious performance bottleneck, as discussed later in this paper, owing to its difficulty of parallelization.}
\yyadd{The pdsyngst routine is an improved routine that employs the rank 2k update, instead of pdtrsm in pdsygst.} 
\yyadd{Since rank 2k update is more suitable for parallelization, pdsyngst is expected to outperform pdsygst.} 
\yyadd{We note that pdsyngst requires more working space (memory) than pdsygst and that pdsyngst only supports lower triangular matrices; 
if these requirements are not satisfied, pdsygst is called \yyyadd{instead of} pdsyngst.} 
\yyadd{For more details of differences between pdsygst and pdsyngst, refer \yyyadd{to} Refs.~\cite{SEARS1998,POULSON2013}.} 
All the workflows except  $A2$ are supported in the \lq current' code,
while the workflow $A2$ was added to EigenKernel very recently in a developer version. 
The workflow $A2$ and other workflows with  pdsyngst will appear in a future version.
It should be noted that all the $3 \times 5$ combinations in Fig.~\ref{FIG-WORKFLOW}
are possible in principle but \yyyadd{some of them} \yadd{have not yet been} implemented in the code, 
owing to the limited human resource for \yadd{programming}. 

\begin{table}
\caption{Available workflows for GEP solver in EigenKernel.}
\label{TABLE-WORKFLOW}       
\begin{tabular}{cll}
\hline\noalign{\smallskip}
Workflow & SEP solver & Reducer  \\
\noalign{\smallskip}\hline\noalign{\smallskip}
A & ScaLAPACK  & ScaLAPACK (pdsygst) \\
A2 & ScaLAPACK  & ScaLAPACK (pdsyngst)  \\
B & Eigen\_sx  & ScaLAPACK (pdsygst) \\
C & ScaLAPACK  & ELPA \\
D & ELPA2 & ELPA \\
E & ELPA1 & ELPA \\
F & Eigen\_s  & ELPA \\
G & Eigen\_sx  & ELPA \\
\noalign{\smallskip}\hline
\end{tabular}
\end{table}

\subsection{Features of EigenKernel}

As stated in Introduction, EigenKernel prepares basic functions to assist the user to use the optimal workflow for GEP. First, it provides a unified interface to the GEP solvers. When the SEP solver and the reducer are chosen from different libraries, the conversion of data format and distribution are also performed automatically. Second, it outputs detailed performance data such as the elapsed times of internal routines of the SEP solver and reducer for performance analysis. The data file is written in JSON (\fdel{=}\finfo{removed $=$}JavaScript Object Notation) format. This data file is used by the performance prediction tool to be discussed in Sec.~\ref{SEC-PREDICTION}. 

In addition to these, EigenKernel has additional features so as to satisfy the needs among application researchers:
(I) It is possible to build EigenKernel only with ScaLAPACK. This is because there are supercomputer systems in which ELPA or EigenExa are not installed.
(II) The package contains a mini-application called EigenKernel-app, a stand-alone application \yyyadd{that reads the matrix data from the file and calls EigenKernel to solve the GEP}. This mini-application can be used for real researches, as in Ref.~\cite{HOSHI2016SC16}, if the matrix data are prepared as files in the Matrix Market format. 

It is noted that there is another reducer routine called EigenExa-style reducer that appears in our previous paper \cite{IMACHI-JIT2016} but is no longer supported by EigenKernel. This is mainly because the code (KMATH\_EIGEN\_GEV)\ffadd{\cite{EIGENGEV}} 
\fdel{\footnote{http://www.r-ccs.riken.jp/labs/lpnctrt/en/projects/kmath-eigen-gev/}}
requires EigenExa but is not compatible with EigenExa 2.4p1. Since this reducer uses the eigendecomposition of the matrix $B$, instead of the Cholesky decomposition of Eq.~(\ref{EQ-CHOL}), its elapsed time is always larger than that of the SEP solver. Such a reducer routine is not efficient, at least judging from the benchmark data of the SEP solvers reported in the previous paper \cite{IMACHI-JIT2016} and in this paper.

\section{Scalability analysis on Oakforest-PACS \label{SEC-PERFORMANCE-ANA}}

In this section, we demonstrate how EigenKernel can be used for performance analysis. We first show the benchmark data of various GEP workflows on Oakforest-PACS obtained using EigenKernel. Then we analyze the performance bottleneck found in one of the workflows with the help of the detailed performance data output by EigenKernel.

\subsection{Benchmarks data for different workflows \label{SEC-BENCH-OFP-DIFF-WORKFLOW}}

The benchmark test was carried out on Oakforest-PACS \fdel{(OFP)}, 
so as to compare the elapsed time among the workflows. 
\fdel{OFP}\fadd{Oakforest-PACS} is \ffadd{a}\fdel{the} massively parallel supercomputer \fdel{installed 
at the Information Technology Center of The University of Tokyo}
\fadd{operated by Joint Center for Advanced High Performance Computing (JCAHPC)~\cite{JCAHPC}}. 
It consists of $P_{\rm max} \equiv$ 8,208 computational nodes connected 
by the Intel Omni-Path network. 
Each node has an Intel Xeon Phi 7250 many-core processor with 3TFLOPS of peak performance. 
Thus, the aggregate peak performance of the system is more than 25PFLOPS.
In EigenKernel, 
the MPI/OpenMP hybrid parallelism is used and 
the number of the used nodes is denoted as $P$.
The number of MPI  processes per node or that of OMP threads per node
is denoted as $n_{\rm MPI/node}$ or $n_{\rm OMP/node}$, respectively.
The present benchmark test was carried out 
with $(n_{\rm MPI/node}, n_{\rm OMP/node}) =(1,64)$.
The present benchmark is limited to those within the regular job classes and
the maximum number of nodes in the benchmark
is $P = P_{\rm \yadd{quarter}} \equiv 2,048$, a \yadd{quarter} of the whole system, because
a job with $P_{\rm \yadd{quarter}}$ nodes is the largest resource available for the regular job classes. 
A benchmark with up to the full system $(P_{\rm \yadd{quarter}} <  P \le P_{\rm max})$ 
is beyond the regular job classes and is planed  in a near future. 
The test numerical problem is \lq VCNT90000' \fdel{that}\fadd{, which} appears in 
ELSES matrix library \fdel{(\verb|http://www.elses.jp/matrix/|)}\fadd{\cite{ELSESMATRIX}}.
The matrix size of the problem is $M=90,000$.
The problem comes from the simulation of a vibrating carbon nanotube (VCNT) 
calculated by ELSES \fdel{(\verb|http://www.elses.jp/|)}
\cite{ELSES,ELSESWEB}, a quantum nanomaterial simulator.
The matrices of $A$ and $B$ in Eq.~(\ref{EQ-GEP-ORG}) were generated 
with an {\it ab initio}-based modeled (tight-binding) electronic-state theory \cite{CERDA}.  

The calculations were carried out by the workflows of $A, A2, D, E, F$ and $G$.
The results of the workflows $B$ and $C$ can be estimated from those of the other workflows,
since the SEP solver and the reducer in the two workflows appear
among other workflows. 
The above discussion implies that the two workflows are not efficient.

\begin{figure*}
  \includegraphics[width=0.99\textwidth]{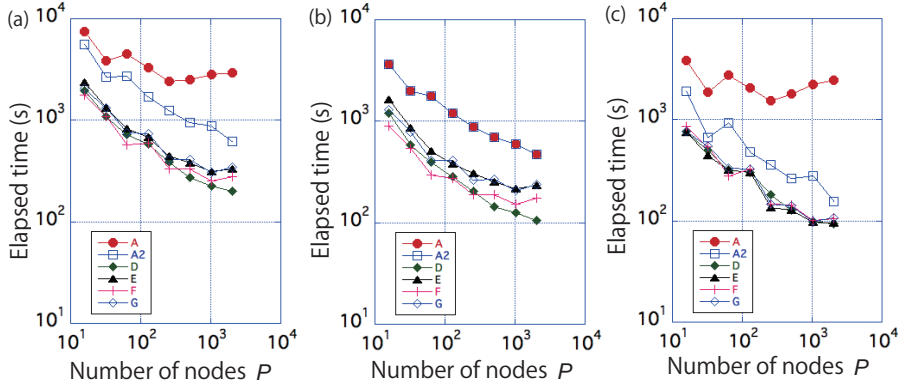}
\caption{
Benchmark on Oakforest-PACS.
The matrix size of the problem is $M=90,000$. 
The computation was carried out with $P=16, 32, 64, 128, 256, 512, 1024, 2048$ nodes
in the workflows of $A$(circle), $A2$(square), $D$(filled diamond), $E$(triangle), $F$(cross), $G$(open diamond).
The elapsed times of (a) for the whole GEP solver, 
(b) for the SEP solver and (c) for the reducer are plotted. 
}
\label{FIG-SCALAPACK-BENCH-DETAIL}
\end{figure*}

The benchmark data is summarized in Fig.~\ref{FIG-SCALAPACK-BENCH-DETAIL}.
The total elapsed time $T(P)$ is plotted in 
Fig.~\ref{FIG-SCALAPACK-BENCH-DETAIL}(a)
and that of the SEP solver $T_{\rm SEP}(P)$ or the reducer 
$T_{\rm red}(P) \equiv T(P) - T_{\rm SEP}(P)$
is plotted in Fig.~\ref{FIG-SCALAPACK-BENCH-DETAIL}(b) or (c), respectively.
Several points are discussed here; 
(I) The optimal workflow seems to be $D$ or $F$ as far as 
among the benchmark data $(16 \le P \le P_{\rm \yadd{quarter}}=2048)$. 
(II) All the workflows except the workflow of $A$ show 
a strong scaling property  in Fig.~\ref{FIG-SCALAPACK-BENCH-DETAIL}(a),
because the elapsed time decreases with the number of nodes $P$. 
Figs.~\ref{FIG-SCALAPACK-BENCH-DETAIL}(b) and (c) indicates that
the bottleneck of the workflow $A$ stems not \yyyadd{from} the SEP solver but \yyyadd{from} the reducer. 
The bottleneck disappears in the workflow $A2$, in which
the routine of pdsygst is replaced by pdsyngst.
(III) The ELPA-style reducer is used in the workflows of $C, D, E, F, G$. 
Among them, 
the workflows $F$ and $G$ are hybrid workflows 
between ELPA and EigenExa and  require 
the conversion process of distributed data,
since the distributed data format is different between ELPA and EigenExa
\cite{IMACHI-JIT2016}. 
The data conversion process 
does not dominate the elapsed time, 
as discussed in Ref.~\cite{IMACHI-JIT2016}.
(IV) We found that 
the same routine gave different elapsed times in the present benchmark.
For example, 
the ELPA-style reducer  is used both in the workflows of $D$ and $E$ but
the elapsed time $T_{\rm red}$ with $P=256$ nodes is significantly different;
The time is $T_{\rm red}$ = 182 s or 137 s, in the workflow of $D$ or $E$, respectively.
The difference stems from the time for the transformation of eigenvectors by Eq.~(\ref{EQ-BACK}), 
since the time is $T_{\rm trans-vec}$ = 69 s or 23 s in the workflow of $D$ or $E$, respectively.
The same phenomenon was observed also in the workflow of $F$ and $G$ with $P$=64 nodes,
since $(T_{\rm red},T_{\rm trans-vec})$=(277 s, 57 s) or (334 s, 114 s)
in the workflow of $F$ or $G$, respectively. 
Here we should remember that
\yadd{even if we use the same number of nodes $P$, the parallel computation time $T=T(P)$ can differ from one run to another,}
since the geometry of \yadd{the} used nodes \yadd{may not be} equivalent. 
Therefore,
the benchmark test for multiple runs with the same number of used nodes
should be carried out in \yadd{a} near future. 
(V) The algorithm has several tuning parameters,
such as $n_{\rm MPI/node}, n_{\rm OMP/node}$,
though these parameters are fixed in the present benchmark.
A more extensive benchmark with different values of the tuning parameters 
is one of possible \yyyadd{areas} of investigation in \yadd{the} future for faster computations.

\subsection{Detailed performance analysis of the pure ScaLAPACK workflows}

\begin{figure*}
  \includegraphics[width=0.95\textwidth]{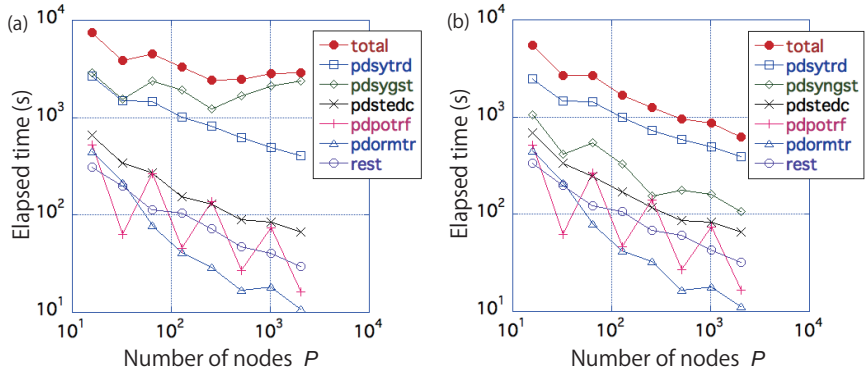}
\caption{
Benchmark on Oakforest-PACS with the workflows (a) $A$ and (b) $A2$.
The matrix size of the problem is $M=90,000$. 
The computation was carried out with $P=16, 32, 64, 128, 256, 512, 1024, 2048$ nodes. 
The graph shows the elapsed times for the total GEP solver (filled circle), 
pdsytrd (square), pdsygst in the workflow $A$ or pdsyngst in the workflow $A2$ (diamond), pdstedc (cross), pdpotrf (plus), pdormtr(triangle)
and the rest part (open circle).
}
\label{FIG-DATA-OFP-SCALAPACK}
\end{figure*}

Detailed performance data
are shown in Fig.~\ref{FIG-DATA-OFP-SCALAPACK} 
for the two pure ScaLAPACK workflows $A$ and $A2$.  
In the workflow $A$, 
the total elapsed time, denoted as $T$, is decomposed into six terms;
\fdel{The}\fadd{the} five terms are those for the ScaLAPACK routines of 
pdsytrd, pdsygst, pdstedc, pdormtr and pdotrf.
The elapsed times for these routines are denoted as
$T^{\rm (pdsygst)}(P)$, $T^{\rm (pdsytrd)}(P)$, $T^{\rm (pdstedc)}(P)$, 
$T^{\rm (pdotrf)}(P)$ and $T^{\rm (pdormtr)}(P)$, respectively.
The elapsed time for the rest part is defined as 
$T^{\rm (rest)} \equiv T- T^{\rm (pdsygst)} - T^{\rm (pdsytrd)} - T^{\rm (pdstedc)}
- T^{\rm (pdotrf)} - T^{\rm (pdormtr)}$. 
In the workflow $A2$, the same definitions are used, 
except the point that pdsygst is replaced by pdsyngst.
These timing data are output by EigenKernel automatically in JSON format.

Figure \ref{FIG-DATA-OFP-SCALAPACK} indicates
that the performance bottleneck of the workflow $A$ 
is caused by the routine of pdsygst,
in which the reduced matrix $A'$ is generated
by Eq.~(\ref{EQ-GEN-MAT-A2}). 
A possible cause for the low scalability of pdsygst is the algorithm used in it, 
which exploits the symmetry of the resulting matrix $A^{\prime}$ 
to reduce the computational cost \cite{WILKINSON}. 
Although this is optimal on sequential machines, 
it brings about some data dependency and 
can cause a scalability problem on massively parallel machines. 
The workflow $A2$ uses pdsyngst instead of pdsygst and 
improves the scalability, as expected 
\yyyadd{from} the last paragraph of Sec.~\ref{SEC-GEP-HYB}.
We should remember that the ELPA-style reducer is \yyyadd{the} best 
among the benchmark in Fig.~\ref{FIG-SCALAPACK-BENCH-DETAIL}(c), 
since the ELPA-style reducer forms the inverse matrix $U^{-1}$ explicitly 
and computes Eq.~(\ref{EQ-GEN-MAT-A2}) directly using matrix multiplication \cite{ELPA}. 
While this algorithm is computationally more expensive,
it has a larger degree of parallelism and can be more suited for massively parallel machines.

In principle, the performance analysis such as given here could be done manually. However, it requires the user to insert a timer into every internal routine and output the measured data in some organized format. Since EigenKernel takes care of these kinds of troublesome tasks, it makes performance analysis easier for non-expert users.
\ffadd{In addition, since }\fadd{performance data obtained in practical computation is sometimes valuable for finding performance issues that rarely appears in development process, 
this feature can contribute to the co-design of software.}

%
\section{Performance prediction  \label{SEC-PREDICTION}}

\subsection{The concept}

\begin{figure*}
  \includegraphics[width=0.5\textwidth]{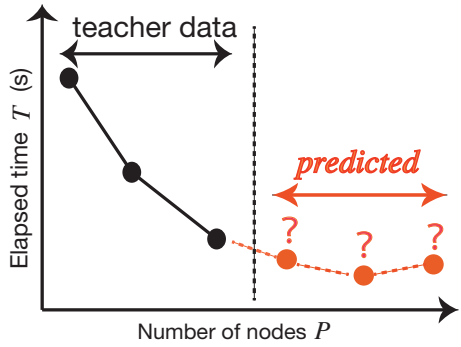}
\caption{Schematic figure of performance prediction, in which
the elapsed time of a routine $T$ is written as the function of the number of processor nodes $P$
($T \equiv T(P)$). 
The figure illustrates the performance extrapolation 
that gives a typical behavior with a minimum. }
\label{FIG_PREDICT_SCHEMATIC}
\end{figure*}

This section proposes to use Bayesian inference as a tool for performance prediction, in which the elapsed time is predicted from teacher data or existing benchmark data. The importance of performance modeling and prediction has long been recognized by library developers. In fact, in a classical paper published in 1996 \cite{Dackland96}, Dackland and K{\aa}gstr{\"o}m write, ``we suggest that any library of routines for scalable high performance computer systems should also include a corresponding library of performance models''. However, there have been few parallel matrix libraries equipped with performance models so far. The performance prediction method to be described in this section will be incorporated in the future version of EigenKernel and will form one of the distinctive features of the middleware.

Performance prediction can be used in a variety of ways.
Supercomputer users \yadd{are required to} prepare
a computation plan that requires estimated elapsed time,
but it is difficult to predict the elapsed time 
from hardware specifications, 
such as peak performance, memory and network \fdel{band widths}\fadd{bandwidths} and communication latency.
The performance prediction realizes high usability,  
since it can predict the elapsed time without requiring huge benchmark data. 
Moreover, the performance prediction enables
an auto-optimization (autotuning) function, 
which selects the optimal workflow in EigenKernel automatically 
given the target machine and the problem size.
Such high usability is crucial,  
for example, in electronic state calculation codes,
because the codes are used not only among \yadd{theorists}
but also among experimentalists and industrial researchers
who are not familiar with HPC techniques.

The present paper focuses 
not only on performance interpolation but also 
on extrapolation, which predicts the elapsed time
at a larger number of nodes from the data
at a smaller number of nodes. This is shown schematically 
in Fig.~\ref{FIG_PREDICT_SCHEMATIC}. 
An important issue in the extrapolation is to predict the speed-up \lq saturation', 
or the phenomenon that the elapsed time may have a minimum, 
as shown in Fig.~\ref{FIG_PREDICT_SCHEMATIC}. 
The extrapolation technique is important, 
since we have only few opportunities
to use the ultimately large computer resources,
like the whole system of the K computer
or Oakforest-PACS. 
A reliable extrapolation technique
will encourage real researchers to use large resources.

\subsection{Performance models}

The performance prediction will be realized,
when a reliable performance model is constructed
for each routine, so as to reflect the algorithm and architecture properly.
The present paper, as an early-stage research, 
proposes three simple models for 
the elapsed time of the $j$-th routine $T^{(j)}$
as the function of the number of nodes (the degrees of freedom in MPI parallelism) $P$;
\fdel{The}\fadd{the} first proposed model is called generic three-parameter model and is expressed as  
\begin{eqnarray}
T^{(j)}(P) &=& T_1^{(j)}(P) + T_2^{(j)}(P) + T_3^{(j)}(P) 
\label{EQ-PERF-MODEL1} \\
T_1^{(j)}(P)  &\equiv& \frac{c_1^{(j)}}{P} 
\label{EQ-PERF-MODEL-TERM1} \\
T_2^{(j)}(P)  &\equiv& c_2^{(j)} 
\label{EQ-PERF-MODEL-TERM2} \\
T_3^{(j)}(P)  &\equiv& c_3^{(j)} \log P
\label{EQ-PERF-MODEL-TERM3} 
\end{eqnarray}
with the three fitting parameters of $\{ c_i^{(j)} \}_{i=1,2,3}$. 
The terms of $T_1^{(j)}$ or $T_2^{(j)}$ stand for
the time in ideal strong scaling or 
in non-parallel computations, respectively.
The model of $T^{(j)} =T_1^{(j)} + T_2^{(j)}$
is known as Amdahl's relation \cite{AMDAHL}.
The  term of $T_3^{(j)}$ stands for the time of MPI communications. 
The logarithmic function was chosen as a reasonable one,
\fadd{since the main communication pattern required in dense matrix computations is categorized into \textit{collective} communication 
(e.g. {\tt MPI\_Allreduce} for calculating inner product of a vector), and such communication routine is often implemented 
as a sequence of point-to-point communications along with a binary tree, whose total cost is proportional to $\log_2 P$ \cite{Pacheko96}.}

The generic three-parameter model in Eq.~(\ref{EQ-PERF-MODEL1}) 
can give, unlike Amdahl's relation, 
the minimum schematically shown in Fig.~\ref{FIG_PREDICT_SCHEMATIC}.  
It is noted that the real MPI communication time is not measured to determine the parameter $c_3^{(j)}$,
since it would require detailed modification of the source code or the use of special profilers.
Rather, all the parameters $\{ c_i^{(j)} \}_{i=1,2,3}$ are estimated simultaneously from the total elapsed time $T^{(j)}$ using Bayesian inference, as will be explained later.

The second proposed model is called generic five-parameter model and is expressed as
\begin{eqnarray}
T^{(j)}(P) &=& T_1^{(j)}(P) + T_2^{(j)}(P) + T_3^{(j)}(P) + T_4^{(j)}(P) + T_5^{(j)}(P)
\label{EQ-PERF-MODEL2} \\
T_4^{(j)}(P)  &\equiv& \frac{c_4^{(j)}}{P^2} 
\label{EQ-PERF-MODEL-TERM4} \\
T_5^{(j)}(P)  &\equiv& c_5^{(j)} \frac{\log P}{\sqrt{P}},
\label{EQ-PERF-MODEL-TERM5} 
\end{eqnarray}
with the five fitting parameters of $\{ c_i^{(j)} \}_{i=1,2,3,4,5}$. 
The term of $T_4^{(j)}(\propto P^{-2})$ is responsible for the \lq super-linear' behavior 
in which the time decays faster than $T_1^{(j)} (\propto P^{-1})$.
The super-linear behavior can be seen in several benchmarks data \cite{Ristov16}.
The term of $T_5^{(j)}(\propto \log P / \sqrt{P})$ expresses the time of MPI communications for matrix computation;
\fdel{When}\fadd{when} performing matrix operations on a 2-D scattered $M \times M$ matrix, 
the size of the submatrix allocated to each node is $(M/\sqrt{P})\times(M/\sqrt{P})$. 
Thus, the communication volume to send a row or column of the submatrix is $M/\sqrt{P}$. 
By taking into account the binary tree \fdel{network}\fadd{based collective communication} and multiplying $\log P$, we obtain the term of $T_5^{(j)}(P)$.
The term decays slower than $T_1^{(j)} (\propto P^{-1})$.

The third proposed model results
when  the MPI communication term of $T^{(j)}_3$ in Eq.~({\ref{EQ-PERF-MODEL1})
is replaced by a linear function; 
\begin{eqnarray}
T^{(j)}(P) &=& T_1^{(j)}(P) + T_2^{(j)}(P) + \tilde{T}_3^{(j)}(P) 
\label{EQ-PERF-MODEL3} \\
\tilde{T}_3^{(j)}(P)  &\equiv& \tilde{c}_3^{(j)} P.
\label{EQ-PERF-MODEL-TERM3B} 
\end{eqnarray}
The model is called linear-communication-term model.
We should say that
this model is fictitious, 
because no architecture or algorithm 
used in real research gives a linear term 
in MPI communication, as far as we know. 
The fictitious three-parameter model of Eq.~(\ref{EQ-PERF-MODEL3})
was proposed so as to be compared with the other two models. 
\fadd{Other models for MPI routines are proposed~\cite{Grbovic07,Hoefler10}, 
and comparison with these models will be also informative, which is one of our future works.}

\subsection{Parameter estimation by the Markov Chain Monte Carlo procedure} \label{BAYESEAN}

In this paper, the model parameters are estimated by 
Bayesian inference with the Markov Chain Monte Carlo (MCMC) iterative procedure
and the uncertainty is included in predicted values. 
Here the uncertainty is formulated by the normal distribution, as usual. 
The result appears as the posterior probability distribution of the elapsed time $T$. 
Hereafter, 
the median is denoted as $T_{\rm med}$ and 
the upper and lower limits 
of 95 \% Highest Posterior Density (HPD) interval are denoted
as $T_{\rm up-lim}$ and $T_{\rm low-lim}$, respectively. 
The predicted value appears both in the median value 
of $T_{\rm med}$ and the interval of $[T_{\rm low-lim}, T_{\rm up-lim}]$.

The MCMC procedure was realized by Python with the Bayesian inference module of PyMC ver. 2.36.
The  method is standard and the use of PyMC is not crucial. 
The MCMC procedure was carried out under the preknowledge that
each term of $\{ T_i^{(j)} \}_{i}$ is a fraction of the elapsed time and therefore
each parameter of $\{ c_i^{(j)} \}_{i}$ should be non-negative $(T_i^{(j)} \ge 0, c_i^{(j)} \ge 0)$. 
The details of the method are explained briefly here; 
(I) 
The parameters of $\{ c_i^{(j)} \}_{i}$ are considered to have uncertainty and 
are expressed as probability distributions. 
The prior distribution should be chosen and 
the posterior distribution will be obtained by Bayesian inference.
The prior distribution of the parameters of $c_i^{(j)}$
is set to the uniform distribution in the interval of $[0, c_{i{\rm (lim)}}^{(j)}]$,
where $c_{i{\rm (lim)}}^{(j)}$ is an input.
The upper limit of $c_{i{\rm (lim)}}^{(j)}$ should be chosen so that
the posterior probability distribution is so localized 
in the region of $c_i^{(j)} \ll c_{i{\rm (lim)}}^{(j)}$.
The values of $c_{i{\rm (lim)}}^{(j)}$ depend on problem
and will appear in the next subsection with results.
(II) 
The present Bayesian inference was carried out 
for the logscale variables $(x, y)  \equiv (\log P, \log T)$, 
instead of the original variable of $(P, T)$. 
The prediction on the logscale variables means that
the uncertainty in the normal distribution appears on the logscale variable $y$. 
When the original variables are used,
the width of the 95 \% HPD interval ($|T_{\rm up-lim}-T_{\rm low-lim}|$) 
is on the same order among different nodes and 
is much larger than the median value $T_{\rm med}$ 
for data with a large number of nodes ($|T_{\rm up-lim}-T_{\rm low-lim}| \gg T_{\rm med}$). 
We thought of the use of the logscale variables, 
since we discuss the benchmark data on the logscale variables
as Fig.\ref{FIG-SCALAPACK-BENCH-DETAIL}. 
Another choice of the transformed variables may be a possible future issue.
(III) The uncertainty in the normal distribution is characterized by the standard deviation $\sigma^{(j)}$.
The parameter $\sigma^{(j)}$ is also treated as a probability distribution and
its prior distribution is set to be the uniform one in the interval
of $[0, \sigma_{\rm limit}]$ with a given value of the upper bound $\sigma_{\rm limit}^{(j)}=0.5$.

The MCMC procedure consumed 
only one or a couple of minute(s) by \yyyadd{a} note PC with the teacher data of existing benchmarks.
The histogram of Monte Carlo sample data are 
obtained for the parameters of $\{ c_i^{(j)} \}_i, \sigma^{(j)}$ and the elapsed time of $T^{(j)}(P)$ and
form approximate probability distributions for each quantity.
The MCMC iterative procedure was carried out for each routine independently 
and the iteration number is set to be $n_{\rm MC} = 10^5$. 
In the prediction procedure of the $j$-th routine,
each iterative step 
gives the set of parameter values of 
$\{ c_i^{(j)} \}_i, \sigma^{(j)}$ and
the elapsed time of $T^{(j)}(P)$, 
according to the selected model. 
We discarded the data 
in the first $n_{\rm MC}^{(\rm early)} = n_{\rm MC}/2$ steps,
since the Markov chain has not converged to the stationary distribution during such early steps.
After that,
the sampling data were picked out with an interval of $n_{\rm interval}=10$ steps. 
The number of the sampling data, therefore, is 
$n_{\rm sample} = (n_{\rm MC}-n_{\rm MC}^{(\rm early)})/n_{\rm interval}=5000$.
Hereafter the index among the sample data is denoted 
by $k$ ($\equiv 1,2,\ldots,n_{\rm sample}$). 
The $k$-th sample data consist of the set of the values
of $\{ c_i^{(j)} \}_i, \sigma^{(j)}$ and $T^{(j)}(P)$
and these values are denoted as 
$\{ c_i^{(j)[k]} \}_i, \sigma^{(j)[k]}$ and $T^{(j)[k]}(P)$, respectively. 
The sampling data set of $\{ T^{(j)[k]}(P) \}_{k=1,...,n_{\rm sample}}$
form the histogram or the probability distribution for $T^{(j)}(P)$.
The probability distributions for the model parameters of $\{ c_i^{(j)} \}_i$ 
are obtained in the same manner  and will appear later in this section.
The sample data for the total elapsed time is given
by the sum of those over the routines; 
\begin{eqnarray}
T^{[k]}(P)  = \sum_j T^{(j)[k]}(P).
\label{EQ-TOTAL-TIME}
\end{eqnarray}
Finally, the median $T_{\rm med}(P)$ and the upper and lower limits
of 95 \% HPD interval, ($T_{\rm up-lim}(P), T_{\rm low-lim}(P)$),  are obtained
from the histogram of $\{ T^{[k]}(P) \}_k$.

\begin{table}
\caption{Measured elapsed times in seconds 
for the matrix problem of \lq VCNT22500' solved by the workflow $A$ on the K computer.
The elapsed times are measured as a function of the number of nodes $P$
for the total solver time $T(P)$ and the six routines of 
$T^{\rm (pdsytrd)}(P)$, $T^{\rm (pdsygst)}(P)$, $T^{\rm (pdstedc)}(P)$, 
$T^{\rm (pdormtr)}(P)$, $T^{\rm (pdotrf)}(P)$, $T^{\rm (rest)}(P)$
}
\label{TABLE-DATA-KEI}       
\begin{tabular}{cccccccc}
\hline\noalign{\smallskip}
\# nodes & total & pdsytrd & pdsygst  & pdstedc & pdormtr & pdotrf &  rest \\ 
\noalign{\smallskip}\hline\noalign{\smallskip}
4 & 1872.7 & 1562.2 & 61.589 & 58.132 & 122.91 & 20.679 & 47.190 \\
16 & 240.82 & 129.09 & 37.012  & 21.341 & 24.624  & 8.0851 & 20.670 \\
64  & 103.18 & 44.494 & 24.584  & 9.9665 & 7.1271 & 3.3122 & 13.692  \\
256  & 63.029 & 21.325 & 20.509 & 5.8159 & 3.4131 & 2.2474 & 9.7189 \\
1024  & 55.592 & 17.524 & 17.242  & 6.1105 & 2.6946 & 3.1462 & 8.8753 \\
4096  & 70.459 & 20.479 & 21.169 & 6.7494 & 3.9326 & 7.9400 & 10.189 \\
10000  & 140.89 & 29.003 & 49.870 & 17.714 & 9.9534 & 19.817 & 14.536 \\
\noalign{\smallskip}\hline
\end{tabular}
\end{table}

\subsection{Results}

The prediction was carried out on the K computer
for the matrix problem of \lq VCNT22500',
which appears in ELSES matrix library \cite{ELSESMATRIX}.
The matrix size is $M=22,500$. 
The workflow $A$, pure ScaLAPACK workflow,  was used
with the numbers of nodes for $P=4, 16, 64, 256, 1024, 4096, 10000$.
The elapsed times were measured for 
the total elapsed time of $T(P)$ and 
the six routines of 
$T^{\rm (pdsygst)}(P)$, $T^{\rm (pdsytrd)}(P)$, $T^{\rm (pdstedc)}(P)$, 
$T^{\rm (pdotrf)}(P)$, $T^{\rm (pdormtr)}(P)$ and $T^{\rm (rest)}(P)$.
The values are shown in Table \ref{TABLE-DATA-KEI}.
The measured data of the total elapsed time of $T(P)$ shows  
the speed-up \lq saturation' or the phenomenon that
the elapsed time shows a minimum at $P  = 1024$.
The saturation is reasonable from HPC knowledge, 
since the matrix size is on the order of $M=10^4$ and 
efficient parallelization can not be expected for $P \ge 10^3$. 

\begin{figure*}
  \includegraphics[width=1.00\textwidth]{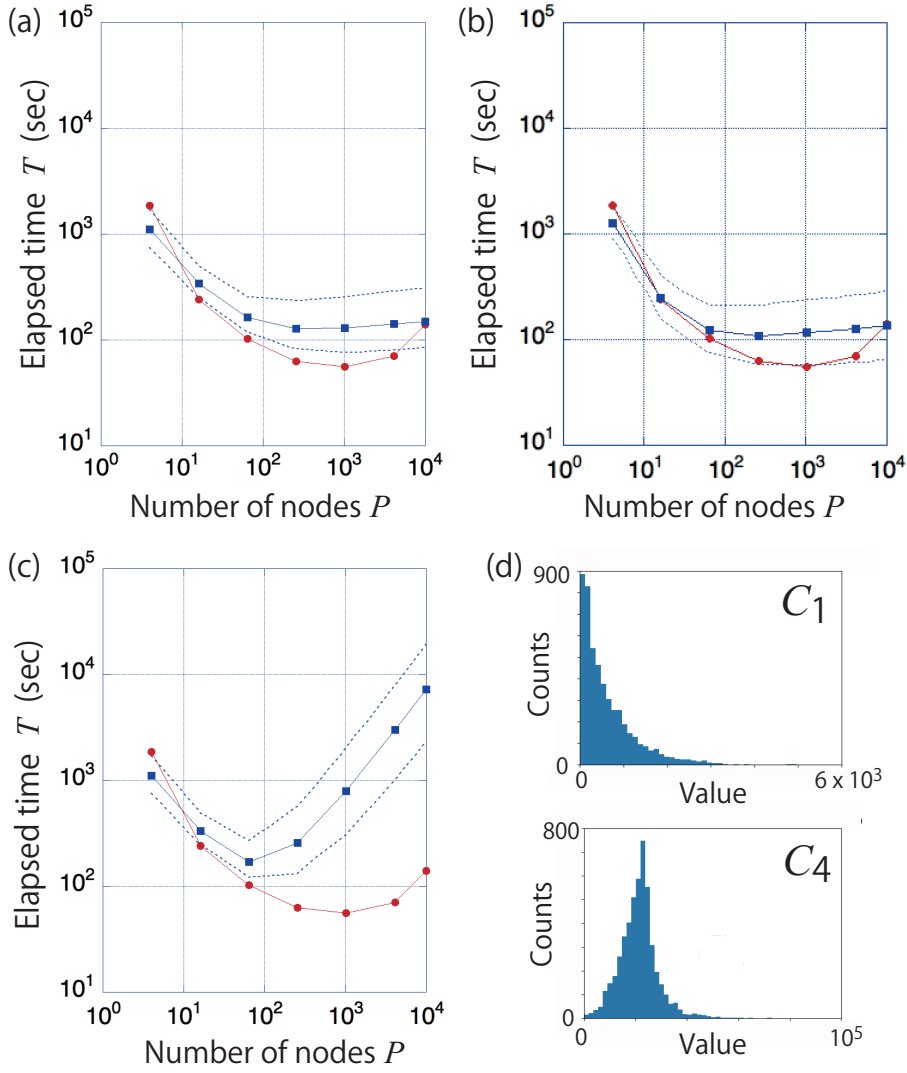}
\caption{Performance prediction with the three models
of (a) the generic three-parameter model, 
(b) the generic five-parameter model, 
and (c) the linear-communication-term model.
The workflow $A$, pure ScaLAPACK workflow,  was used
with the numbers of nodes for $P=4, 16, 64, 256, 1024, 4096, 10000$.
The data is generated for the generalized eigenvalue problem 
of \lq VCNT22500' on the K computer.
The measured elapsed time for the total solver is drawn by circles.
The predicted elapsed time is drawn
by square for the median and
by dashed lines for the upper and lower limits
of 95 \% HPD interval.
The used teacher data are
the elapsed times of the six routines of 
$T^{\rm (pdsytrd)}$, $T^{\rm (pdsygst)}$, $T^{\rm (pdstedc)}$, 
$T^{\rm (pdotrf)}$, $T^{\rm (pdormtr)}$ and $T^{\rm (rest)}$ at $P=4, 16, 64$. 
(d) The posterior probability distribution 
of $c_1^{\rm (pdsytrd)}$ (upper panel) 
and $c_4^{\rm (pdsytrd)}$ (lower panel) 
within the five-parameter model. 
}
\label{FIG-PREDICT}
\end{figure*}

Figure\fadd{s} \ref{FIG-PREDICT}(a)-(c) shows the result of  Bayesian inference,
in which the teacher data is the measured elapsed times 
of the six routines of 
$T^{\rm (pdsygst)}(P)$, $T^{\rm (pdsytrd)}(P)$, $T^{\rm (pdstedc)}(P)$, 
$T^{\rm (pdotrf)}(P)$, $T^{\rm (pdormtr)}(P)$ and $T^{\rm (rest)}(P)$ at $P=4, 16, 64$. 
In the MCMC procedure,
the value of $c_{i{\rm (lim)}}^{(j)}$ is set
to $c_{i{\rm (lim)}}^{(j)} = 10^5$ among all the routines
and the posterior probability distribution satisfies the locality 
of  $c_i^{(j)} \ll c_{i{\rm (lim)}}^{(j)}$.
One can find 
that the generic three-parameter model of Eq.~(\ref{EQ-PERF-MODEL1}) 
and the generic five-parameter model of Eq.~(\ref{EQ-PERF-MODEL2}) commonly
predict the speed-up saturation successfully at $P = 256 \sim 1024$,
while the linear-communication-term model does not. 
Examples of the posterior probability distribution are shown in
Fig.~\ref{FIG-PREDICT}(d), 
for $c_1^{\rm (pdsytrd)}$ and $c_4^{\rm (pdsytrd)}$ in the five-parameter model.

\begin{figure*}
  \includegraphics[width=1.00\textwidth]{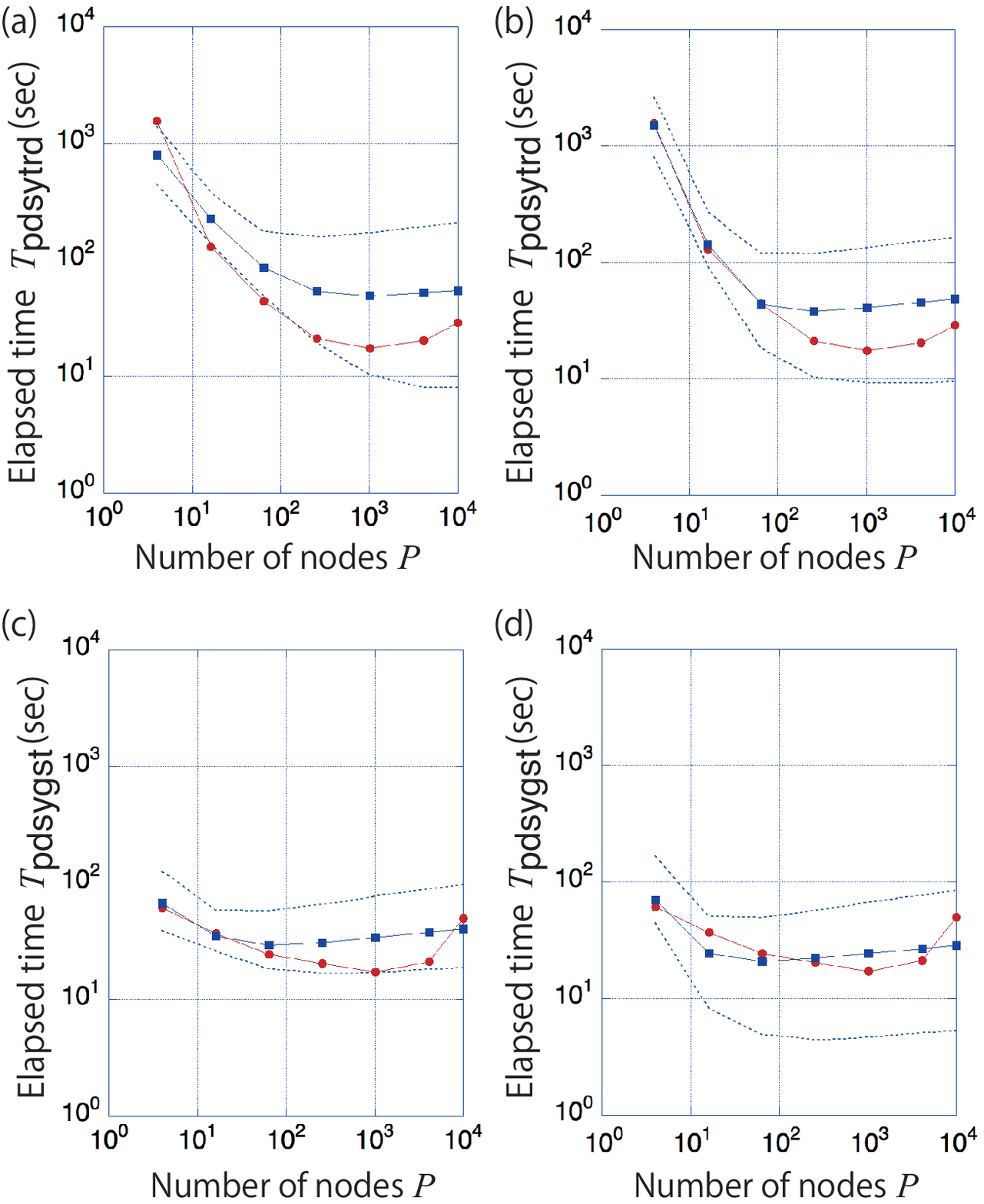}
\caption{Detailed analysis of the performance prediction in Fig.~\ref{FIG-PREDICT}.
The performance prediction for $T^{\rm (pdsytrd)}$ is shown 
by (a) the generic three-parameter model and 
(b) the generic five-parameter model.
The performance prediction for $T^{\rm (pdsygst)}$ is shown 
by (c) the generic three-parameter model and 
(d) the generic five-parameter model.
}
\label{FIG-PREDICT-DETAIL}
\end{figure*}

Figure \ref{FIG-PREDICT-DETAIL} shows 
the detailed analysis of the performance prediction in Fig.~\ref{FIG-PREDICT}.
Here the performance prediction of $T^{\rm (pdsytrd)}$ and $T^{\rm (pdsygst)}$
is focused on, since the total time is dominated by these two routines,
as shown in Table \ref{TABLE-DATA-KEI}. 
The elapsed time of $T^{\rm (pdsytrd)}$
is predicted in Fig\fadd{s}.~\ref{FIG-PREDICT-DETAIL}(a) and (b) 
by the generic three-parameter model and the generic five-parameter model,
respectively. 
The importance of the five-parameter model can be understood,
when one see the probability distribution of $c_1^{\rm (pdsytrd)}$
and $c_4^{\rm (pdsytrd)}$ in Fig.~\ref{FIG-PREDICT}(d).  
Since the probability distribution of $c_4^{\rm (pdsytrd)}$ has a peak at  
$c_4^{\rm (pdsytrd)} \approx 2.3 \times 10^4 $, 
the term of $T_4^{(j)}(P)$  in Eq.~(\ref{EQ-PERF-MODEL-TERM4}), the super-linear term, 
should be important. 
The contribution at $P=10$, for example, is estimated to be
$T_4(P=10) = c_4/P^2 \approx (2.3 \times 10^4) / 10^2 \approx 2 \times 10^2$. 
The above observation is consistent with the fact that
the measured elapsed time shows the super-linear behavior 
between $P=4,16$ 
($T(P=4)/T(P=16)$=(1872.7 sec)/(240.82 sec) $\approx$ 7.78)
and the generic five-parameter model reproduces
the teacher data, the data at $P$=4, 16, 64, 
better than the generic three-parameter model. 
The elapsed time of $T^{\rm (pdsygst)}$
is predicted in Fig\fadd{s}.~\ref{FIG-PREDICT-DETAIL}(c) and (d) 
by the generic three-parameter model and the generic five-parameter model,
respectively. 
The prediction by 
the generic five-parameter model seems to be better
than that by the three-parameter model. 
Unlike $T^{\rm (pdsytrd)}$, 
$T^{\rm (pdsygst)}$ is a case in a poor strong scaling, 
since the ideal scaling behavior $(T \propto 1/P)$ or
the super-linear behavior $(T \propto 1/P^2)$ can not be seen 
even at the small numbers of used nodes $(P=4,16)$.


\begin{figure*}
  \includegraphics[width=1.00\textwidth]{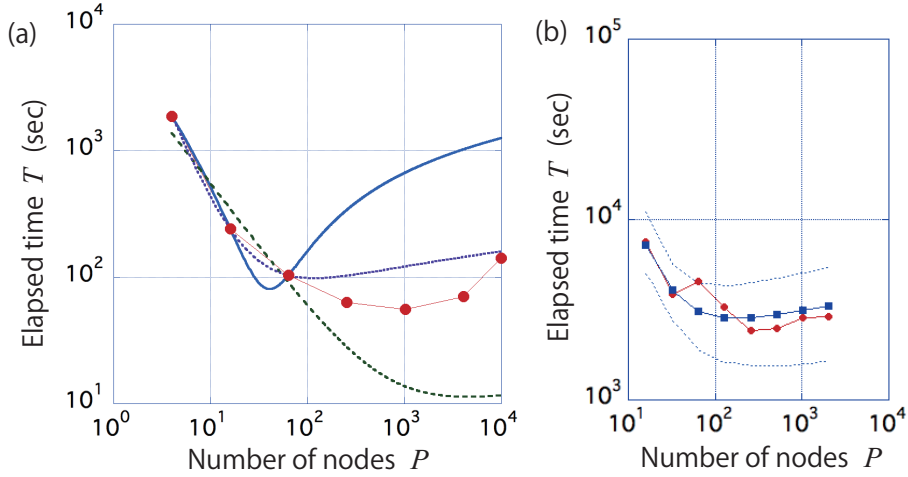}
\caption{(a) Performance prediction  by least squares methods
of \lq VCNT22500' on the K computer. 
The red circles indicate the measured elapsed times 
for $P=4, 16, 64, 256, 1024, 4096, 10000$,
while the other curves are determined by the least squares methods
with the teacher data at $P=4, 16, 64$; 
The bold solid line is determined without any constraint
by the three-parameter model.  
The dashed line is determined with non-zero constraint 
by the three-parameter model.  
The dotted line is determined with non-zero constraint 
by the five-parameter model.  
In the fitting procedure by the five-parameter model, 
the initial values of the iterative procedure
are chosen to be the result by the three-parameter model. 
(b) Performance prediction for the generalized eigenvalue problem 
of \lq VCNT90000' on Oakforest-PACS. 
The generic five-parameter model is used. 
The red circles indicate the measured elapsed times 
for $P$=16, 32, 64, 128, 256, 512, 1024 and 2048.
The predicted elapsed time is drawn
by square for the median and
by dashed lines for the upper and lower limits
of 95 \% HPD interval.
The teacher data are the data at $P=$16, 32, 64, 128. 
}
\label{FIG-CONVENTIONAL-FIT}
\end{figure*}

\section{Discussion on performance prediction}
\label{DISCUSSION}

\subsection{Comparison with conventional least squares method}

In this subsection, the present MCMC method is compared with 
the conventional least squares methods,
so as to clarify the properties of the present method. 
The least squares methods have been applied to the performance modeling of numerical libraries. 
In fact, most of the recent studies on performance modeling \cite{Peise12,Reisert17,Fukaya15,Fukaya18}
rely on the least squares methods to fit the model parameters.

The fitting results by three types of least squares methods are shown in Fig.~\ref{FIG-CONVENTIONAL-FIT}(a). 
The data for VCNT22500 on the  K computer is used, as in Fig.~\ref{FIG-PREDICT}.
The total elapsed time of $T(P)$ is fitted 
with the teacher data at $P=4, 16, 64$. 
The bold solid line is the fitted curve 
by the least squares method, without any constraint, 
in the generic three-parameter model. 
The fitting procedure determines the parameters
as $(c_1,c_2,c_3) = (c_1^{\rm (LSQ0)},c_2^{\rm (LSQ0)},c_3^{\rm (LSQ0)}) 
\equiv (1.06 \times 10^{4}, -1.14 \times 10^{3}, 2.60 \times 10^{2})$.
The fitted curve reproduces the teacher data, the data at $P=4, 16, 64$, exactly, 
but deviates severely from the measured values at $P \ge 256$.
The fitted value of $c_2$ is negative,
since the method ignores the preknowledge
of the non-negative constraint $(c_2 \ge 0)$. 
The dashed and dotted lines are the fitted curves 
by the least squares methods under the non-negative constraint on the fitting parameters of $\{ c_i \}_i$.
The fitting procedure was carried out by the module lsqnonneg  in MATLAB. 
The dashed line is fitted with the three-parameter model and 
the fitted values are 
$(c_1,c_2,c_3) = (c_1^{\rm (LSQ1)},c_2^{\rm (LSQ1)},c_3^{\rm (LSQ1)}) \equiv (5.47 \times 10^3, 3.20 \times 10^{-10}, 2.77)$. 
The dotted line is fitted with the five-parameter model and 
the fitted values are 
$(c_1,c_2,c_3,c_4,c_5) = (c_1^{\rm (LSQ2)},c_2^{\rm (LSQ2)},c_3^{\rm (LSQ2)},c_4^{\rm (LSQ2)},c_5^{\rm (LSQ2)}) 
\equiv (1.65 \times 10^3, 5.64 \times 10^{-2}, 17.2, 2.30 \times 10^{4}, 4.99 \times 10^{-3})$. 
The dotted curve is comparable to 
the median values in the MCMC method of Fig.~\ref{FIG-PREDICT}(b).
We found, however, that 
the fitting procedure with the five-parameter model is problematic, 
because the fitting problem is underdetermined, having smaller number of teacher data ($n_{\rm teacher}=3$) than that of fitting parameters ($n_{\rm param}=5$), and therefore the objective function can have multiple local minima.
The fitting procedure is iterative and 
we chose the initial values as
$(c_1,c_2,c_3,c_4,c_5) = 
(c_1^{\rm (LSQ1)},c_2^{\rm (LSQ1)},c_3^{\rm (LSQ1)},0, 0)$ 
in the case of Fig.~\ref{FIG-CONVENTIONAL-FIT}(a).
We found that several other choices of the initial values fail to converge.

The above numerical experiment reflects
the general difference between the MCMC method and the least square methods. 
In general, the MCMC method has the following three properties; 
(i) The non-negative constraint on the elapsed time ($T>0$) can be imposed as preknowledge. 
(ii) The uncertainty or error can be taken into 
account both for the teacher data and the predicted data. 
The uncertainty of the predicted data appears as the HPD interval, for example,   
in Fig.~\ref{FIG-PREDICT}. 
(iii) The iterative procedure is guaranteed to converge to the unique posterior distribution. 
The least square method without any constraint
does not have any of the above three properties,
as is exemplified by the bold solid curve case of Fig.~\ref{FIG-CONVENTIONAL-FIT}(a). 
The least square method with non-negative constraint has only the property (i),
which is reflected in the dashed and dotted curve cases of Fig.~\ref{FIG-CONVENTIONAL-FIT}(a). 
The fitting procedures do not contain the uncertainty, because of the lack of the property (ii). 
Instead of the property (iii), the least squares method gives the parameter values that 
attain one of the local minima of the least squares objective function,
which can be problematic, as explained at the end of  the previous paragraph.

In addition, it is noted 
that the least squares method by the three-parameter model 
without constraint can be realized
in the framework the MCMC method,
if the parameters of
$\{ c_i \}_i$ are set to be the interval of $[-c_{\rm lim}, c_{\rm lim}]$
with $c_{\rm lim} =10^5$
and that for  $\sigma$ is set to be the interval of $[0, \sigma_{\rm lim}]$
with $\sigma_{\rm lim} = 10^{-5}$.
The above prior distribution means that
the non-negative condition $(c_i \ge 0)$ is ignored and
the method is required to reproduce the teacher data exactly
($\sigma_{\rm lim} \approx 0$).
In this case, the MCMC procedure was carried out
for the  variables $(x, T)  \equiv (\log P, T)$, unlike in Sec.~\ref{BAYESEAN},
because
the non-negative condition is not imposed on $T$ and
we cannot use $y \equiv \log T$ as a variable.
We confirmed the above statement
by the MCMC procedure.
As results,
the median is located at
$(c_1,c_2,c_3) = (c_1^{\rm (LSM)},c_2^{\rm (LSM)},c_3^{\rm (LSM)})$
and the width of
the 95 \% HPD interval is tiny ($10^{-3}$ or less)
for each coefficient.

\subsection{Possible extension of the present models}

Although the generic five-parameter model seems to be the best
among the three proposed models,
its theoretical extension is possible for more flexible models.
Figure \ref{FIG-CONVENTIONAL-FIT}(b) shows the performance prediction by the five-parameter model 
for the benchmark data 
by the workflow $A$ in Fig.~\ref{FIG-DATA-OFP-SCALAPACK},
a data for the problem of \lq VCNT90000' on Oakforest-PACS. 
The data at $P=16, 32, 64, 128$ are the teacher data. 
\hadd{In the present case, the upper limit is set
to $c_{i{\rm (lim)}}^{(j)} = 10^5$ for $i=1,2,3,5$ and
$c_{4{\rm (lim)}}^{(j)} = 10^7$ and 
the posterior probability distribution satisfies 
the locality of $c_i^{(j)} \ll c_{i{\rm (lim)}}^{(j)}$.}
The predicted data fails to reproduce 
the local maximum at $P=64$ in the measured data,
since the model can have only one (local) minimum and no (local) maximum. 
If one would like to overcome the above limitation, 
a candidate for a flexible and useful model may be one with case classification; 
\begin{eqnarray}
T(P) = 
\left\{ 
\begin{array}{ll}
T_{\rm model}^{(\alpha)}(P) & \quad P < P_{\rm c} \\
T_{\rm model}^{(\beta)}(P) & \quad P > P_{\rm c} \\
\end{array}
\right.,
\end{eqnarray}
where $T_{\rm model}^{(\alpha)}(P)$ and 
$T_{\rm model}^{(\beta)}(P)$ are considered to be independent models and
the threshold number of $P_{\rm c}$ is also a fitting parameter. 
A model with case classification will be fruitful from the algorithm and architecture viewpoints.
An example is the case where the target numerical routine switches the algorithms according to the number of used nodes.
Another example is the case where
the nodes in a rack are tightly connected and 
parallel computation within these nodes is quite efficient. 
From the application viewpoint, however,
the prediction in Fig.~\ref{FIG-CONVENTIONAL-FIT}(b) is still meaningful, 
since the extrapolation implies that 
an efficient parallel computation cannot be expected at $P \ge 128$.

\subsection{Discussions on methodologies and future aspect}

This subsection is devoted to discussions on methodologies and future aspects.

\hadd{
(I) 
The proper values of $c_{i{\rm (lim)}}^{(j)}$ 
should be chosen in each problem and here we propose a way to set the values automatically. 
The possible maximum value of $c_{i}^{(j)}$ appears, when
the elapsed time of the $j$-th routine is governed only by the $i$-th term of the given model
($T^{(j)}(P) \approx T_i(P)^{(j)} $). 
We consider, for example, the case in which the first term is dominant
($T^{(j)}(P) \approx c_{1}^{(j)}/P $) and the possible maximum value of 
$c_{1}^{(j)}$ is given by $c_{1}^{(j)} \approx PT^{(j)}(P)$. 
Therefore the limit of $c_{1{\rm (lim)}}^{(j)}$ can be chosen to 
be $c_{1{\rm (lim)}}^{(j)}  \ge PT^{(j)}(P)$ among the teacher data of $(P, T^{(j)}(P))$. 
The locality of $c_{1}^{(j)} \ll c_{1{\rm (lim)}}^{(j)}$ should be checked
for the posterior probability distribution.
We plan to use the method in a future version of our code.
}

(II) 
The elapsed times of parallel computation may be different among multiple runs,
as discussed in the last paragraph of Sec.~\ref{SEC-BENCH-OFP-DIFF-WORKFLOW}. 
Such uncertainty of the measured elapsed time can be treated in Bayesian inference by, for example, setting the parameter $\sigma_{\rm lim}^{(j)}$ appropriately based on the sample variance of the multiple measured data or other preknowledge. In the present paper, however, the optimal choice of $\sigma_{\rm lim}^{(j)}$ is not discussed and is left as a future work.

(III) 
It is noted that ATMathCoreLib, an autotuning tool \cite{ATMATHCORELIB1,ATMATHCORELIB2}, 
also uses Bayesian inference for performance prediction. However, it is different from our approach in two aspects. First, it uses Bayesian inference to construct a reliable performance model from noisy observations~\fadd{\cite{Suda10}}. So, more emphasis is put on interpolation than on extrapolation. Second, it assumes normal distribution both for the prior and posterior distributions. This enables the posterior distribution to be calculated analytically without MCMC, but makes it impossible to impose non-negative condition of $c_i\ge 0$.

(IV) 
The present method uses the same generic performance model
for all the routines. A possible next step is to develop a proper model for each routine.
Another possible theoretical extension is performance prediction 
among different problem sizes.
The elapsed time of a  dense matrix solver depends on the matrix size $M$,
as well as on $P$, so it would be desirable to develop a model to express the elapsed time as a function of both the number of nodes 
and the matrix size ($T=T(P,M)$). 
For example, 
the elapsed time of the tridiagonalization routine in EigenExa on the K computer is modeled as a function of both the 
matrix size and the number of nodes~\cite{Fukaya18}. 
In this study, several approaches to performance modeling are compared depending on \ffadd{the information available for modeling}, 
and some of them accurately estimate the elapsed time for a \yyyadd{given} condition (i.e. matrix size and node count). 
The use of such a model will \yyyadd{provide} more fruitful prediction,
in particular, for the extrapolation.

\section{Summary and future outlook}
\label{SUMMARY}

We developed an open-source middleware EigenKernel 
for the generalized eigenvalue problem that realizes high scalability and usability, 
responding to the solid need in large-scale electronic state calculations.
Benchmark results on Oakforest-PACS shows that 
the middleware enables us to construct the optimal hybrid solver
from ScaLAPACK, ELPA and EigenExa routines. 
The detailed performance data provided by EigenKernel 
reveals performance issues without additional effort such as code modification. 
For high usability,
a performance prediction method was proposed based on Bayesian inference
with \yyyadd{the} Markov Chain Monte Carlo procedure.
We found that the method is applicable not only to performance interpolation
but also to extrapolation.
For a future look, we can consider a system that gathers performance data automatically every time users call a library routine. Such a system could be realized through a possible collaboration with the supercomputer administrator and will give a greater prediction power to our performance prediction tool, by providing huge set of teacher data. 
The performance prediction method is general and applicable to any numerical procedure,
if a proper performance model is prepared for each routine.
The present middleware approach 
will form a foundation of the application-algorithm-architecture co-design.

\section*{Acknowledgement}
The authors thank to Toshiyuki Imamura (RIKEN) for the fruitful discussion on EigenExa
and Kengo Nakajima (The University of Tokyo) 
for the fruitful discussion on Oakforest-PACS. 
\yyyadd{We are also grateful to the anonymous reviewers, whose comments helped us improving the quality of this paper.}


\end{document}